\documentclass[journal]{IEEEtran}

\ifCLASSINFOpdf
  \usepackage[pdftex]{graphicx}
\else
  \usepackage[dvips]{graphicx}
\fi
\usepackage[cmex10]{amsmath}
\usepackage[tight,footnotesize]{subfigure}

\usepackage[font=footnotesize]{subfig}
\usepackage{stfloats}
\begin{document}
%
\title{Optical Absorption Characteristics of Silicon Nanowires for Photovoltaic Applications}
%
%
%

\author{Vidur~Parkash,~\IEEEmembership{Student Member,~IEEE,}
        and~Anand~K.~Kulkarni
\thanks{V. Parkash is with the Department
of Electrical and Computer Engineering, Michigan Technological University,Houghton,
MI, 49931 USA e-mail: vparkash@mtu.edu.}
\thanks{A. K. Kulkarni is with the Department
of Electrical and Computer Engineering, Michigan Technological University,Houghton,
MI, 49931 USA e-mail: akkulkar@mtu.edu.}

\thanks{Manuscript received June 1, 2010.}}

\maketitle

\begin{abstract}
Solar cells have generated a lot of interest as a potential source of clean renewable energy 
for the future. However a big bottleneck in wide scale deployment of these energy sources 
remain the low efficiency of these conversion devices. Recently the use of nanostructures and 
the strategy of quantum confinement have been as a general approach towards better charge carrier
generation and capture. In this article we have presented calculations on the optical  characteristics of 
nanowires made out of silicon. Our calculations show these nanowires form excellent optoelectronic
materials and may yield efficient photovoltaic devices.

\end{abstract}


%
\IEEEpeerreviewmaketitle

\section{Introduction}
\IEEEPARstart{I}{n} recent years a lot of innovation in the semiconductor industry has 
enabled solar cell designs with higher conversion efficiencies though the use of inexpensive materials
and processes \cite{luque_1997,lee_2009}. Today III-V semiconductor based solar cells 
can achieve upto 40\% efficiency. Most strategies in improving the efficiencies target
improving optical absorption, and efficient collection of generated photocarriers. In recent times
improving efficiencies through quantum confinement has generated a lot of interest, with the advent of
multijunction cells (MJSCs) and quantum dot based devices \cite{hubbard_2009}. 
However most of the high efficiency devices are still based on exotic III -V hetrostructures, which are
expensive to process and manufacture. Hence a real incentive in realizing cells made out of silicon still
exists due to the existing infrastructure already in place. \\
\par
\noindent
A possible candidate for the next generation of solar cells are silicon nanowires. Silicon nanowires are 1D materials
and there has been interest to incorporate them into photovoltaic cell design due to the wide availability of the material and existing processing infrastructure. The advantage of using nanowires being that, like quantum dots we 
can tune the band gap to a specific part of the solar spectrum, but with the added advantage that carrier transport in the solar cell is not dependent on quantum tunneling as proposed in \cite{conibeer_2008}.
In this paper we have presented the optical properties of silicon nanowires of different dimensions 
evaluated from ab initio calculations. Our results show that the nanowires are direct bandgap materials
and we can alter the energy gap anywhere between 2-5 eV, that gives us a wavelength selectivity from
200 to 600 nm wavelengths. Results on intrinsic absorption and excitonic binding energies are also 
presented later in this article.\\
\par
\noindent
The rest of this paper is organised as follows, Section-II describes the computational methodology followed 
for the ab initio calculations, In Section -III we present our results on bandstructure and optical properties.
like the dielectric function and absorption coefficient. Finally we evaluate how exciton binding energies in
these nanowires vary with the length of the nanowire (corresponds to quantum confinement in the axial direction).


\hfill
 
\section{Computational Methodology}

\begin{figure}[t]
\begin{center}
\subfigure[]{\includegraphics[scale=0.11]{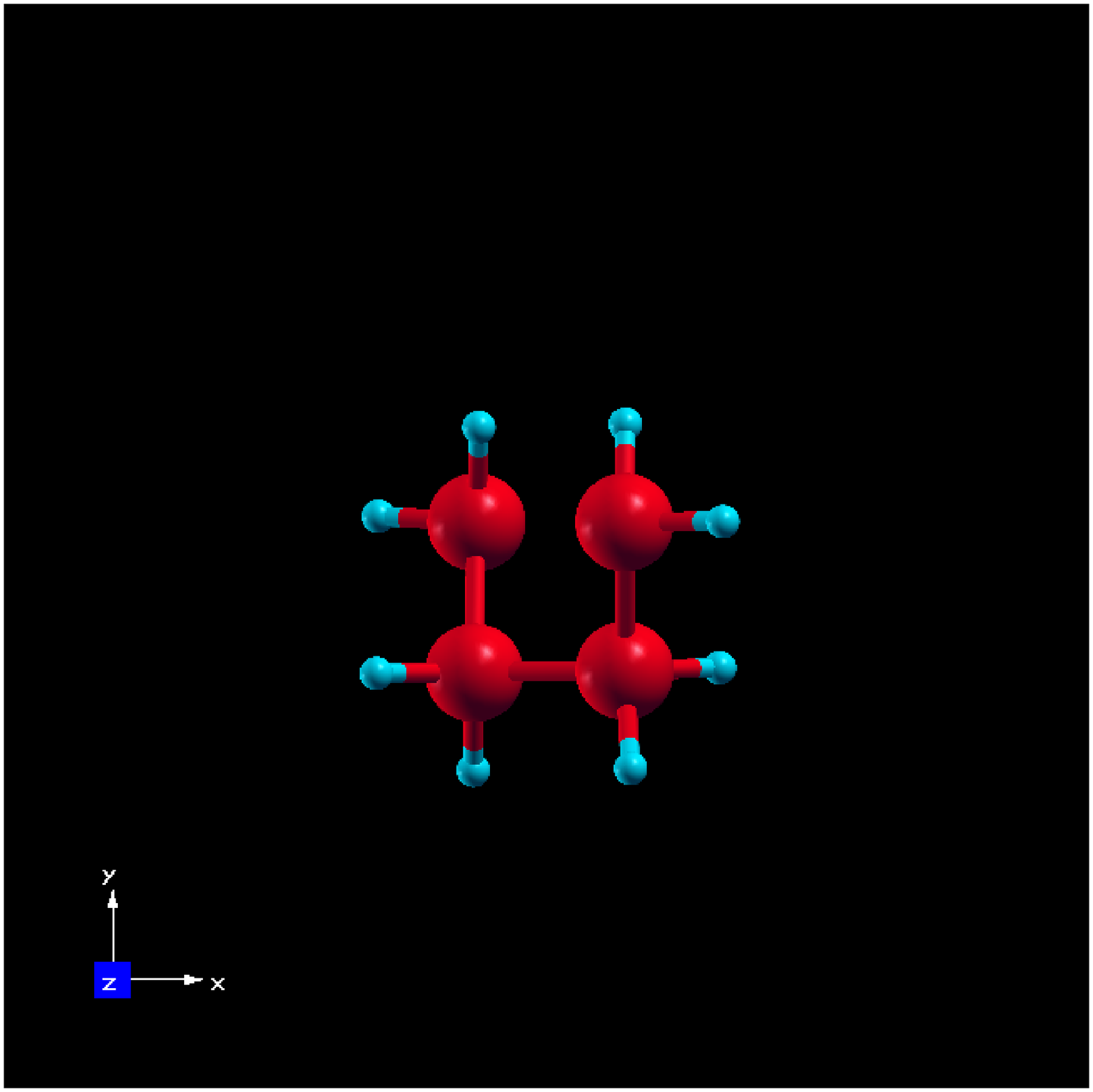}}
\subfigure[]{\includegraphics[scale=0.11]{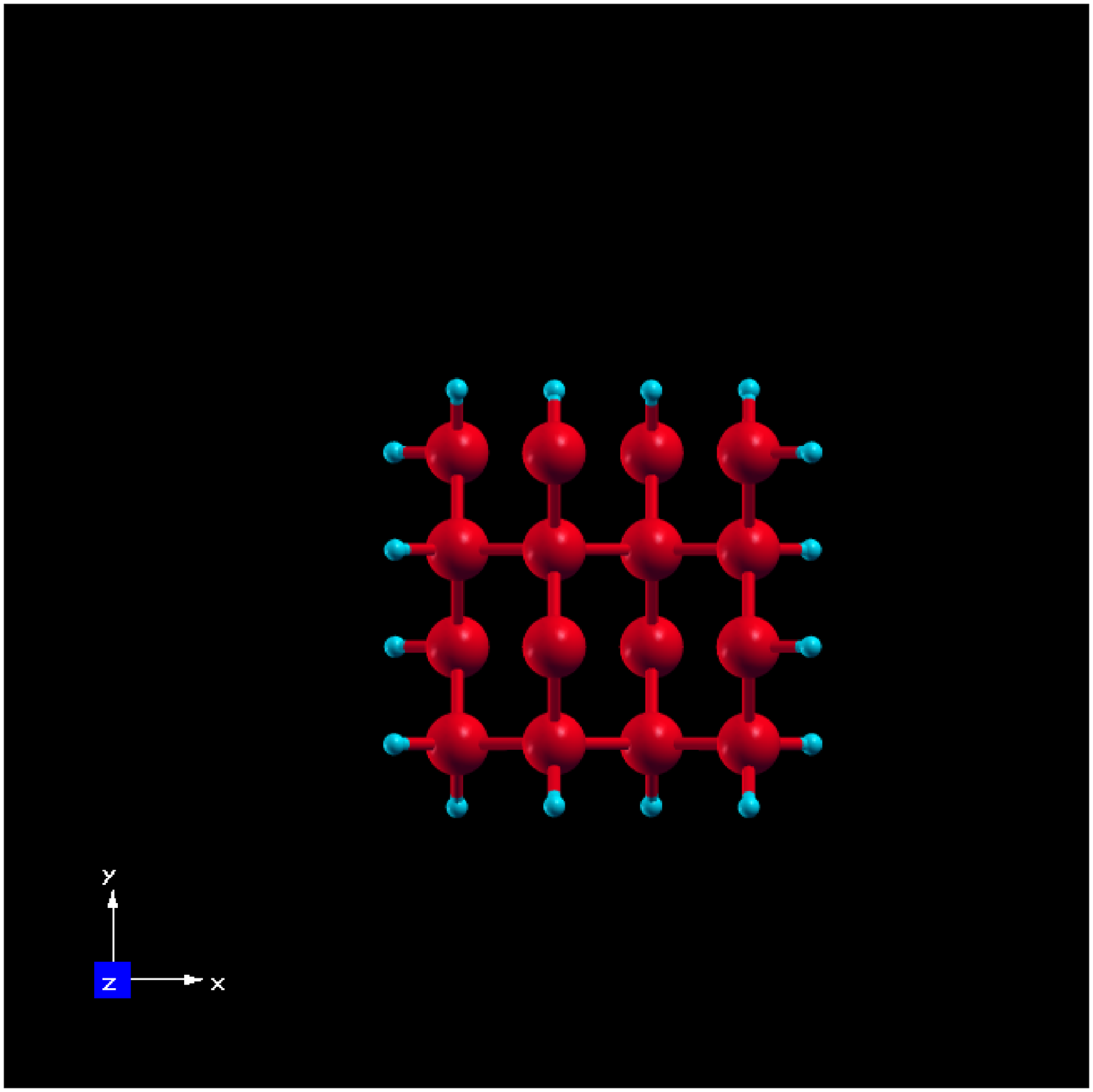}}
\subfigure[]{\includegraphics[scale=0.11]{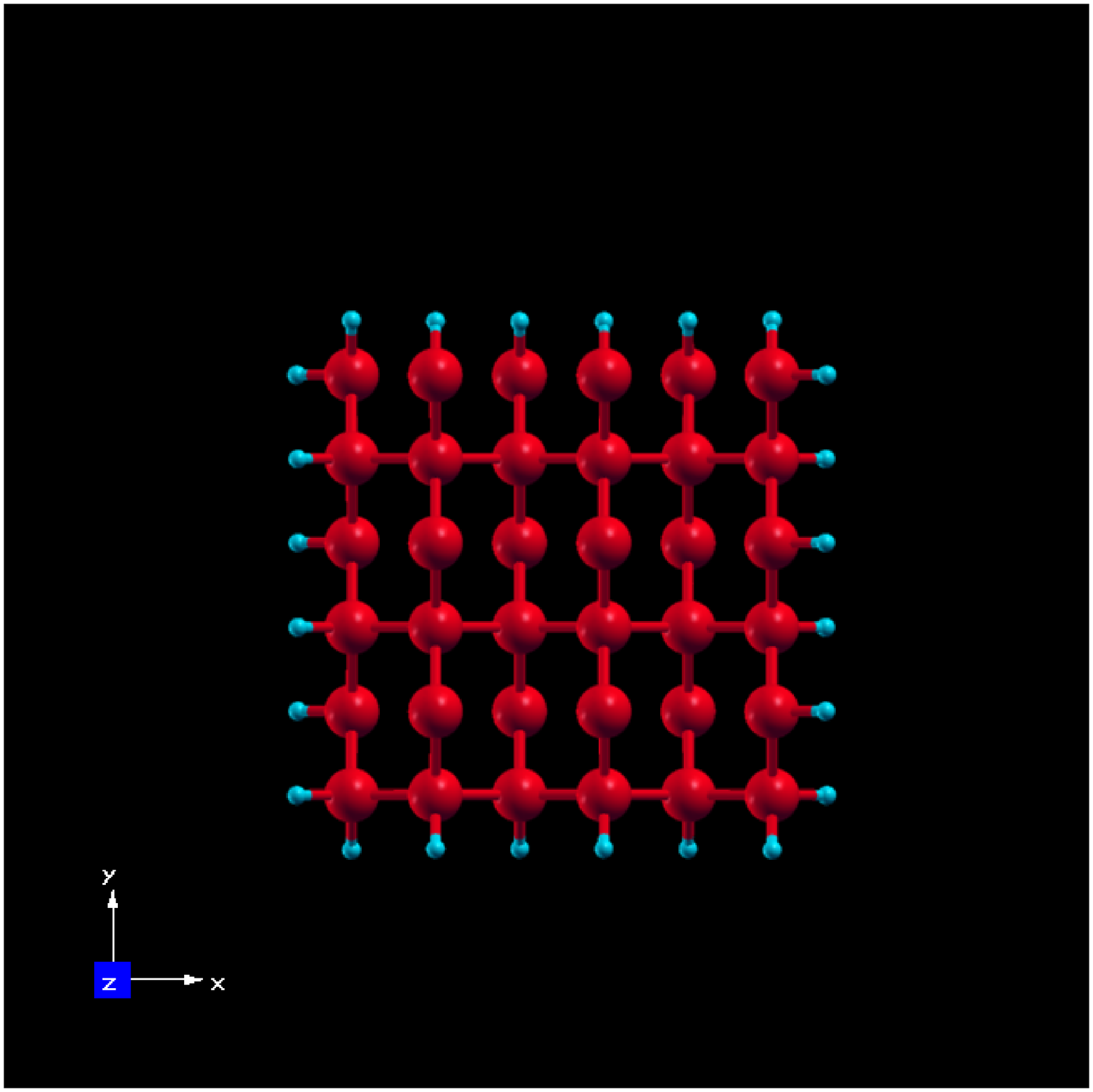}}
\end{center}
\caption{silicon nanowire unit cell geometries seen along the [100] direction. The three 
nanowires are designated (a) Si11,(b) Si22 and (c) Si33 in this paper}
\label{sixx}
\end{figure}

In this section we describe the first principles calculations carried out for the silicon nanowires systems. 
The computations were carried out using a density functional DFT-LDA approach using a parallel binary of the PWScf
distribution. We carried out computations on three nanowires. We consider the nanowires to be oriented along the [100]
plane, each having a square cross section. The dangling bonds on the surface of the nanowire are passivated by
hydrogen terminations. Before performing electronic structure calculations we subjected the unit cells to a geometry 
optimization procedure.The equilibrium lattice relaxed to a unit cell with the Si-Si bond spacing to be 2.324 \AA~and Si-H
 spacing to be 1.5 \AA. The translational lattice constant for the nanowires was 5.43 \AA. 
We have considered three nanowire unit cells in this paper each with a diameter 4.34~\AA, 8.16~\AA~ and 
10.75 \AA. These have been designated as Si11, Si22 and Si33 systems in this paper. The relaxed unit
 cells considered for the calculations presented in this paper are shown in Fig.~\ref{sixx}. The silicon atoms
 in the Si11 unit cell forms the primitive cell for the Si22, Si33 nanowires as well. The Si\textbf{XX}
system is formed by '\textbf{X}' repetitions of this atomic arrangement in the x and y directions. Each of the atoms 
in the Si11 cell occupy four different z-planes at $z=0$, $z= \pm a_0/4$ and $z=a_0/2$.\\
\par
\noindent
The electronic structure of each of the nanowire unit cells was computed self consistently using a plane wave pseudopotential
method. We utilized Van Barth Carr soft potentials for both silicon and hydrogen, with a Perdew Zunger exchange and correlation \cite{perdew_1981}.
The SCF calculations were performed with a grid of 20 uniformly spaced k-vectors within the 1D Brilliouin Zone of the
nanowires. A kinetic energy cutoff of 80 Ry was deemed sufficient for numerical convergence of total energy. For the
computation of optical absorption coefficient $\alpha(\omega)$, a Drude-Lorentz broadening of 0.13 eV has been applied.

\section{Electronic and Optical Characteristics}

\subsection{Bandstructure}
\par
\noindent
The silicon nanowires unlike bulk silicon are direct band gap semiconductors, which would make them excellent
choices for optical applications. Authors in \cite{sanders_1992,miranda_2009,bruno_2007} have also reported direct band gaps and monotonically decreading bandgapes from their
theoretical calculations. 

All the nanowires exhibit exceptionally high effective masses of holes and electrons. For example the Si11 wire
has a almost zero dispersion state on the top of the valence band. Similarly, the lowest conduction bands are also 
relatively heavy compared to their values in the bulk state. This is quite advantageous as it allows for extension of 
carrier lifetime as binding energy of excitons are directly dependent on it (\ref{bulk_ex}). 
We found that the energy gap of the nanowires decrease with increasing width.
 The wide tunability in the band gap of the nanowires allows us to tune
the selectivity of optical absorption of the nanowire. The idea behind this being to match optical absorption
in solar cells to the spectrum of the incoming solar radiation. A second intersting feature of the band structure is 
the weak dispersion of the conduction and valence bands leads to a high effective mass both in the conduction and valence bands.
There is some experimental evidence \cite{cui_2000} who have found out that carrier mobilities in intrinsic nanowires 
is extremely low. Table-\ref{table1} shows the effective masses of the lowest four conduction bands $C1\rightarrow C4$ for all the three systems.
Due to the relatively weak dispersion seen in these wires especially the Si11 and Si22 systems, it is prudent to present a thermally averaged effective mass, i.e. a 
harmonic average over a $KT/q$ energy range from the lowest conduction band energy.



\begin{table}[!t]
\renewcommand{\arraystretch}{1.3}

\caption{Computed Oscillator strengths for direct K=0 transitions}
\label{table_osc}
\centering

\setlength{\tabcolsep}{12pt}
\begin{tabular}{l c c c l l}
\hline
\hline
Syst. & VB & CB & $\Delta E$ (eV) & $m_e^*$ & $f_0$\\
\hline
Si11 & V1 & C1 & 5.165 & 1.44 & 1.48e-5 \\
Si11 & V1 & C2 & 5.272 & 1.92 & 0.0196$^\dagger$\\
Si11 & V1 & C3 & 5.275 & 0.58 & 0.0125$^\dagger$\\
Si11 & V1 & C4 & 5.61 & 0.91 & 0.5951$^\dagger$\\

Si11 & V2 & C1 & 5.318 & 1.44 & 0.000313\\
Si11 & V2 & C2 & 5.425 & 1.92 & 1.63e-6\\
Si11 & V2 & C3 & 5.428 & 0.58 & 0.11\\
Si11 & V2 & C4 & 7.763 & 0.91 & 0.00549\\

Si11 & V3 & C1 & 5.327 & 1.44 & 3.21e-5\\
Si11 & V3 & C2 & 5.434 & 1.92 & 0.1759\\
Si11 & V3 & C3 & 5.437 & 0.58 & 1.28e-4\\
Si11 & V3 & C4 & 5.772 & 0.91 & 3.59e-6\\

Si11 & V4 & C1 & 6.623 & 1.44 & 2.07e-5\\
Si11 & V4 & C2 & 6.73 & 1.92 & 0.0360\\
Si11 & V4 & C3 & 6.733 & 0.58 & 0.0209\\
Si11 & V4 & C4 & 7.068 & 0.91 & 3.59e-6\\
\hline
Si22 & V1 & C1 & 3.497 & 0.71 & 7.07e-7 \\
Si22 & V1 & C2 & 3.574 & 0.35 & 7.73e-7\\
Si22 & V1 & C3 & 3.577 & 0.82 & 0.0260\\
Si22 & V1 & C4 & 3.577 & 0.26 & 1.7e-7\\

Si22 & V2 & C1 & 3.498 & 0.71 & 6.97e-7 \\
Si22 & V2 & C2 & 3.574 & 0.35 & 0.0004\\
Si22 & V2 & C3 & 3.578 & 0.82 & 0\\
Si22 & V2 & C4 & 3.578 & 0.26 & 0.0087\\

Si22 & V3 & C1 & 3.558 & 0.71 & 6.96e-7 \\
Si22 & V3 & C2 & 3.634 & 0.35 & 0.0002\\
Si22 & V3 & C3 & 3.638 & 0.85 & 0.0145\\
Si22 & V3 & C4 & 3.638 & 0.26 & 0.0048\\

Si22 & V4 & C1 & 3.69 & 0.71 & 0.204$^\dagger$ \\
Si22 & V4 & C2 & 3.766 & 0.35 & 9.01e-7\\
Si22 & V4 & C3 & 3.77 & 0.85 & 5.41e-5\\
Si22 & V4 & C4 & 3.77 & 0.26 & 1.81e-5\\
\hline

Si33 & V1 & C1 & 2.042 & 0.58 & 4.23e-10 \\
Si33 & V1 & C2 & 2.105 & 1.7 & 0\\
Si33 & V1 & C3 & 2.116 & 0.9 & 1e-4\\
Si33 & V1 & C4 & 2.116 & 0.81 & 0\\

Si33 & V2 & C1 & 2.044 & 0.58 & 4.3e-10 \\
Si33 & V2 & C2 & 2.107 & 1.7 & 1e-4\\
Si33 & V2 & C3 & 2.118 & 0.9 & 0\\
Si33 & V2 & C4 & 2.118 & 0.81 & 7e-6\\

Si33 & V3 & C1 & 2.179 & 0.58 & 1.13e-6 \\
Si33 & V3 & C2 & 2.242 & 1.7 & 2e-11\\
Si33 & V3 & C3 & 2.253 & 0.9 & 3.78e-5\\
Si33 & V3 & C4 & 2.253 & 0.81 & 9.4e-6\\

Si33 & V4 & C1 & 2.194 & 0.58 & 0.005748$^\dagger$ \\
Si33 & V4 & C2 & 2.257 & 1.7 & 8.89e-9\\
Si33 & V4 & C3 & 2.268 & 0.9 & 2.03e-11\\
Si33 & V4 & C4 & 2.268 & 0.81 & 5.6e-12\\
\hline
\hline
\end{tabular}
$\dagger$ indicates a strong absorption peak in Fig.~\ref{sixxalpha}. Calculations for other transitions are weak and not
discussed in this paper. 
\label{table1}
\end{table}

\begin{figure}
\centering
\includegraphics[scale=0.7]{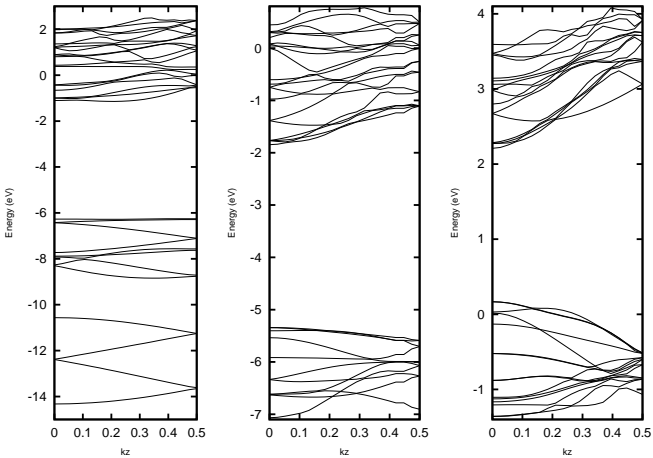}
\caption{1D Bandstructure of silicon nanowire system along the z-direction with $0<k_z<\pi/a_0$ with $a_0 =5.43$~\AA~being
the lattice constant. Band diagrams are shown for (\textit{from left to right}) Si11, Si22 and Si33 systems}
\end{figure}

\begin{figure*}[t]
\centering
\subfigure[]{\includegraphics[scale=0.4]{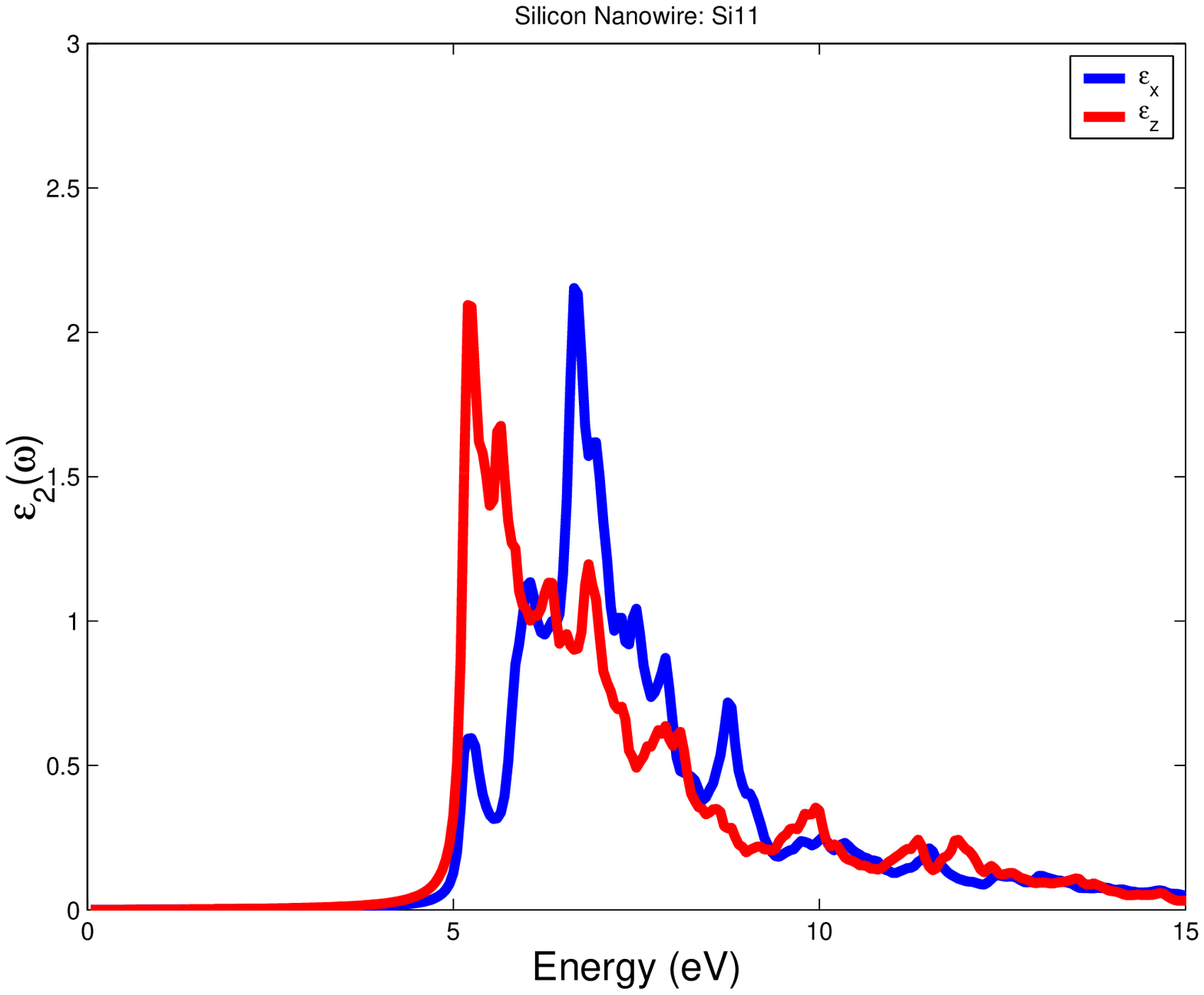}}
\subfigure[]{\includegraphics[scale=0.4]{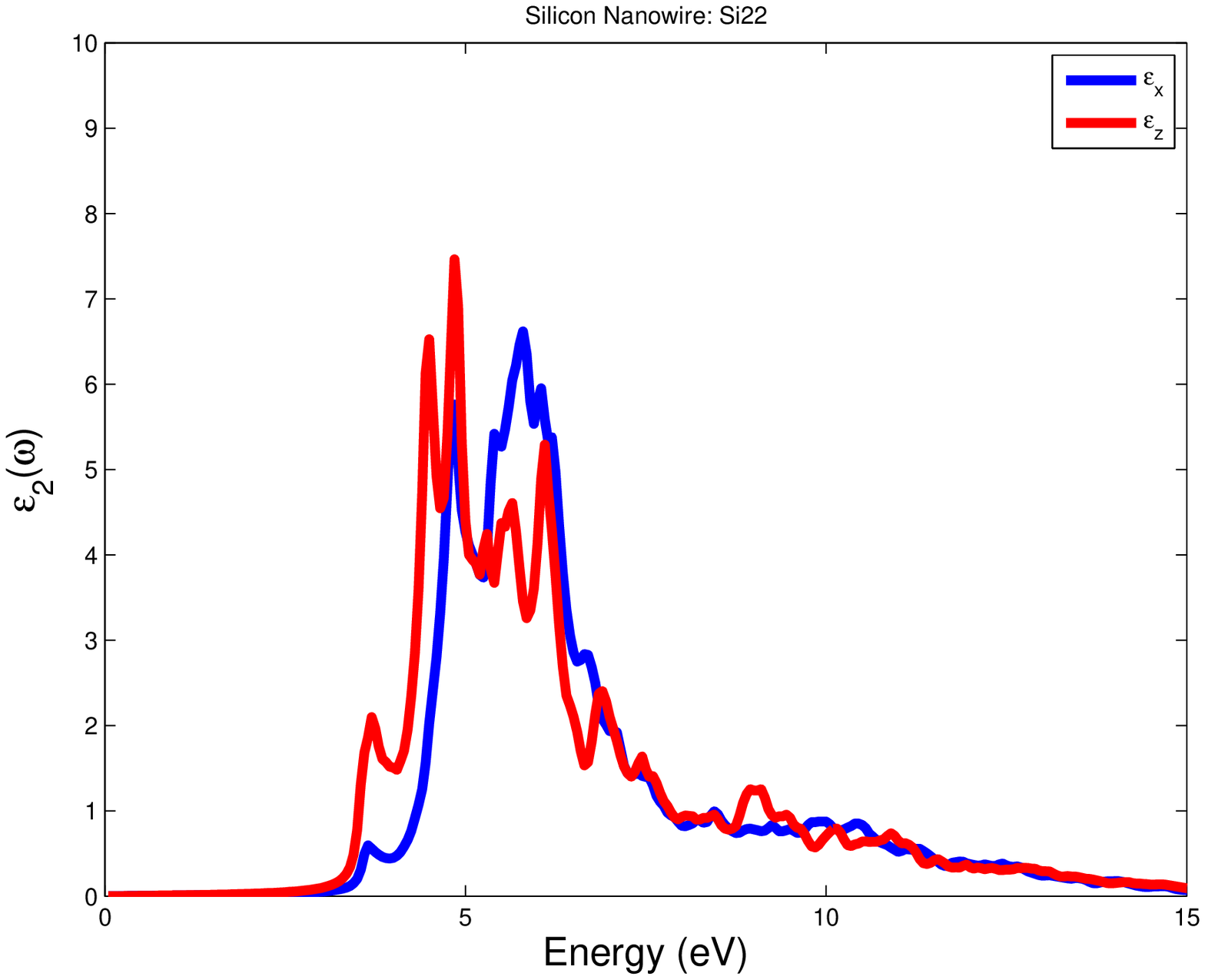}}
\subfigure[]{\includegraphics[scale=0.4]{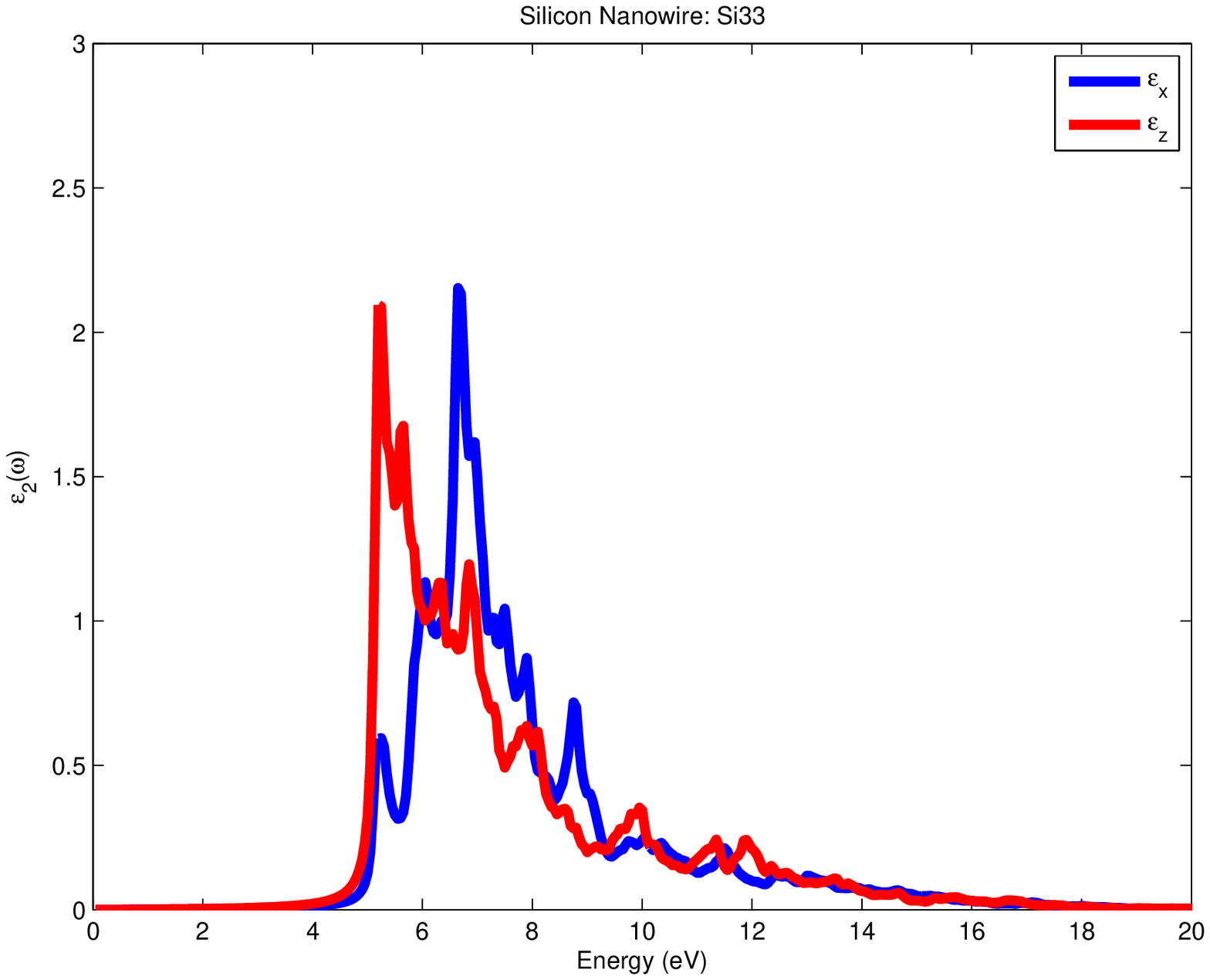}}
\subfigure[]{\includegraphics[scale=0.4]{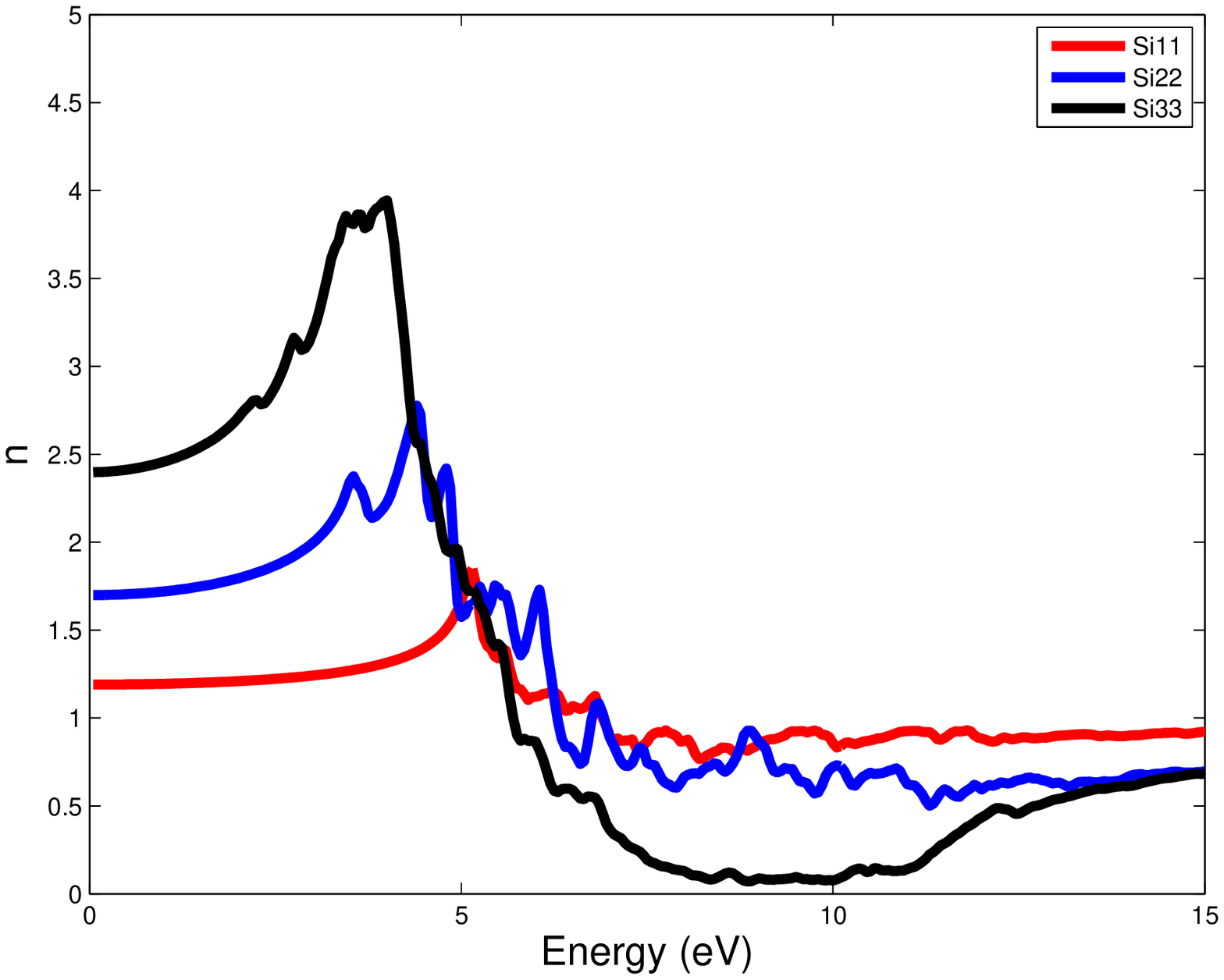}}
\caption{Optical Properties of silicon nanowires (a)-(c) Imaginary part of dielectric function $\epsilon_2(\omega)$ for the
three nanowires shown for both longitudinal and transverse polarizations, (d) Comparison of refractive indices}
\label{fig_e2_n}
\end{figure*}

\subsection{Optical Absorption}
\begin{figure}[!t]
\centering
\includegraphics[width = 75mm, scale=0.3]{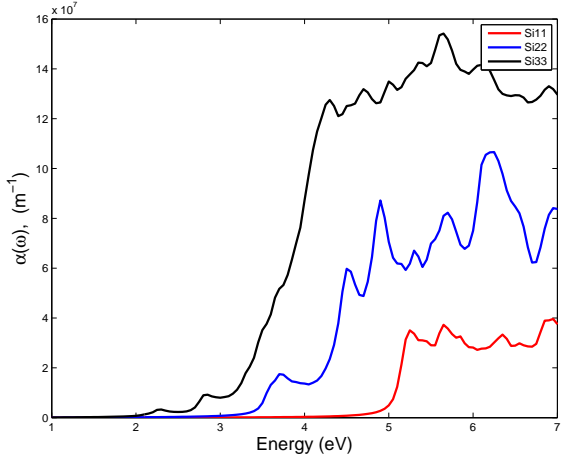}
\caption{Optical Absorption coefficient for the nanowire systems $\alpha(\omega)$ in units of energy (eV).}
\label{sixxalpha}
\end{figure}

\noindent
All the nanowires examined in this paper show features in the absorption spectrum the correspond to excitonic processes. 
A simultaneous comparison of their absorption spectra is shown in Fig.~\ref{sixxalpha}. The lowest excitonic peaks for each of the nanowires occur at
5.25 eV (232 nm), 3.7 eV (335 nm), and 2.3 eV (539 nm) in the increasing order of size. The corresponding optical wavelength
is listed in parentheses. We see that absorption is tunable from the visible region to the near UV portion of the solar 
spectrum. These peaks are also quite prominent in the Si11 wire as compared to the Si33 nanowire due to the a small density of states in the conduction band.
The rippled nature of the absorption spectrum weakens as the size increases. In the Si33 wire, these oscillations are almost washed out and the spectrum
looks similar to that of a direct bandgap material with a exponential absoprtion edge near the conduction band energy \cite{poruba_2004,derlett_1995}.
To ascertain the exact transitions that show up in the absorption spectrum, we computed the dipole matrix element of 
for each of the eigenvalues at $k_z=0$ and computed a quantity called the oscillator strength $f_0$. 
The quantity $f_0$ is proportional to the square of the dipole matrix as

\begin{equation}
f_0=\frac{2m_e}{3\hbar^2}\left(E_2 - E_1\right)\sum_\alpha |\left<\psi_1|R_\alpha|\psi_2\right>|^2
\end{equation}

where the summation is over all possible polarizations. The magnitude of $f_0$ is a direct gauge of the strength of 
absorption at a particular energy. These numbers have been tabulated in Table-\ref{table_osc}. We 
observed that only a few excitations have significant values of $f_0$ appear optically active in the absorption spectrum.
These excitations are also marked in Table-\ref{table_osc}. Other optical properties of these nanowires are
also presented in Fig.~\ref{fig_e2_n}. The strong transition peaks are more clearly visible in the plots for $\epsilon_2(\omega)$.

\subsection{Excitonic Binding energy}

It is known that excitonic processes are helpful in extending lifetimes of the electron hole pairs created by
optical absorption. The radiative lifetimes of an exciton $\tau_r$ is proportional to the binding energy
 $\tau_r \propto E_b^{3/2}$. Hence it is useful to evaluate how change in the length of the nanowire
changes the binding energy of the 1D exciton. To get this dependence we followed the methods of \cite{bastard_1982,gerlach_1998}
solving a two particle Schrodinger equation for an electron-hole system. In this section
we have presented binding energy results for the lowest lying excitonic state in each of the nanowires. The Hamiltonian for
 such a system can be written in the Hartree approximation
\begin{equation}
H = \frac{p_{ze}^2}{2m_e}  +\frac{ p_{zh}^2}{2m_h} - \frac{q^2}{4\pi\epsilon_0\epsilon_r|z_e - z_h|}
\label{ex_h}
\end{equation}

The first two terms are simply the kinetic energy of the electron
and hole respectively and the third term is the screened Coulomb potential. The exciton is assumed to be immersed in
a surrounding medium with an effective dielectric constant $\epsilon_r$. For each of the nanowires 
we used a value computed from the real dielectric constant at zero frequency i.e. $\epsilon_1(0)$.
This Hamiltonian is diagonalized by a two particle wave function of the form

\begin{equation}
\psi = N\cos{\frac{\pi z_e}{L}}\cos{\frac{\pi z_h}{L}}e^{-\frac{z_e-z_h}{\lambda}}
\label{ex_psi}
\end{equation}

where $\lambda$ is a fitting constant and N is a normalizing constant. This test wavefunction was originally applied to 1D quantum wells. This trial wavefunction is acceptable in SiNWs as it is reasonable to conceptually think of a nanowire
as a 1D quantum well with an unsually long width. Equation (\ref{ex_psi}) makes a single assumption that the wavefunction becomes zero at the
ends of the quantum well, which is applicable to the nanowire as well.  To find the binding energy of the exciton we use a variational 
method of minimizing the total energy over different values of $\lambda$. The ground state of the exciton is simply

\begin{equation}
E_b = min_\lambda \frac{\left<\psi|H|\psi\right>}{\left<\psi|\psi\right>} 
\label{energy_min}
\end{equation}

\begin{figure}[t]
\centering
\includegraphics[scale=0.55]{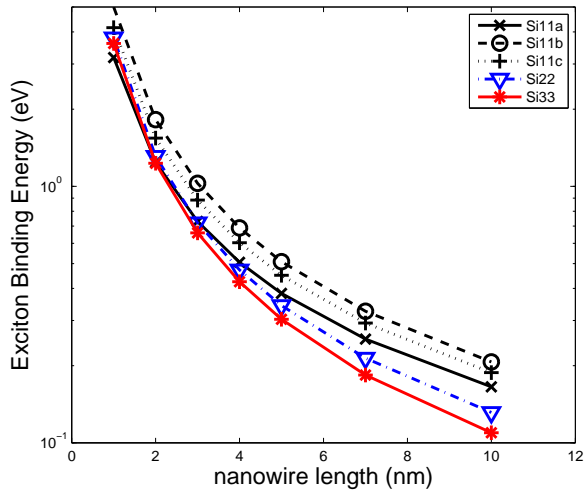}

\caption{1D Exciton binding energy in silicon nanowires as a function of nanowire length. Excitons correspond to direct transitions labeled by a $\dagger$ in
Table-\ref{table1} }
\label{binding_energy}
\end{figure}

The minimized value of $\lambda$ corresponds to the effective Bohr radius of the exciton. The binding
 energies have been plotted as a function of nanowire length in Fig.~\ref{binding_energy}. We see
that the binding energies monotonically increase as the nanowire is made shorter. This dependence
is expected as the stronger confinement causes an increase in the effective mass of the charge carriers. The 
bulk excitonic binding energy is dependent on the reduced mass $\mu$ of the exciton given by the equation

\begin{equation}
E_{bulk} = \frac{\mu q^4}{32 \pi^2\hbar^2 \epsilon_{0}^2\epsilon_{r}^2}
\label{bulk_ex}
\end{equation}

where $\mu=m_e m_h/(m_e + m_h)$. We also see that the excitonic absorption is the strongest in the Si11 system
compared to the shorter excitonic peaks seen in the Si22 and Si33 system (Fig.~\ref{sixxalpha}).Using the method presented above, we
 computed the exciton binding energies for the strongest transitions in the nanowire systems.
There are three strong transitions of interest in the Si11 system and one transition each in the Si22 and Si33 system (see Table-\ref{table_osc}).
 We found that the excitons in the nanowires are
predominantly Mott Wannier type. The typical Bohr radii for the 1D free exciton $\lambda_f$ was found to range between 2.5-5 nm. 
The largest radius being exhibited by the Si33 wire due to a strong screening potential. However confining the exciton in a
short nanowire, with lengths less than $\lambda_f$ causes a large jump in the binding energies of the excitons as the electron
hole pairs are forced to be closer to each other. The binding energies 
have a strong effect on the lifetime of these carriers. The radiative lifetime $\tau$ is related to the oscillator strength
and the exciton binding energy by the relation \cite{perebeinos_2005,seitz_1963}

\begin{equation}
\frac{1}{\tau_r}=\frac{nq^2 E_v^2f_0}{2\pi\epsilon_0 m_e \hbar^2 c^3}
\end{equation}
\\
where $E_v = E_g - E_b$, $n$ is the refractive index of the material and the other symbols have their usual meanings.
We found out that these nanowires exhibit unusually large exciton lifetimes with numbers
of the order of a microsecond. We have summarized our findings for a specimen length of 10 nm (see Table-\ref{table_opt}). 
In the Si11 wire designated \textit{Si11c} the radiative lifetime is about 4 ns ($l=10\textrm~{nm}$ and similar vales of $E_b$), far less than the other two transitions \textit{Si11a} and \textit{Si11b}.
This is simply because of the large oscillator strength of this excitation is also responsible for a quick decay.

\begin{table}[!t]
\renewcommand{\arraystretch}{1.3}

\caption{Radiative Lifetime of lowest optically active excitons}
\label{table_opt}
\centering

\setlength{\tabcolsep}{4pt}
\begin{tabular}{l l l l l l l l l}
\hline
\hline
Syst. & $\epsilon_1(0)$ & n & $E_b$(eV) & $\lambda_{f}$(nm) & $m_e^*$ & $f_0$ & $\tau$ ($\mu s$)\\
\hline
Si11a & 1.5 & 1.2 & 0.1655 & 2.55 & 1.92 & 0.016 & 0.152 \\
Si11b & 1.5 & 1.2 & 0.2068 & 3.457 & 0.58 & $3\times10^{-3}$ & 0.198 \\
Si11c & 1.5 & 1.2 & 0.188 & 3.1 & 0.91 & 0.233 & $4\times10^{-3}$ \\
Si22 & 3 & 1.7 & 0.1314 & 4.216 & 0.71 & 0.051 & 0.022 \\
Si33 & 5 & 2.4 & 0.1095 & 5.08 & 0.58 & $1.43\times10^{-3}$ & 0.955 \\
\hline
\hline
\end{tabular}
\end{table}

\section{Conclusion}
We have presented first principles calculations for silicon nanowires for potential applications
in photovoltaic applications. Results indicate that the nanowires are direct band gap semiconductors
and show excellent wavelength selectivity in the optical region of the solar spectrum. We also identified 
the optically active interband transitions present in these nanowires and used it to calculate 
excitonic effects in the nanowire by solving a two particle Schrodinger equation. Our calculations indicate
that excitons in these nanowires show large radiative lifetimes of the order of several microseconds for 
nanowires as small as 10 nm in length. This number is expected to grow significantly with further confinement
as we move into the quantum dot regime.


%

\ifCLASSOPTIONcaptionsoff
  \newpage
\fi



%
\bibliography{refs}
\bibliographystyle{ieeetr}
%

\begin{IEEEbiographynophoto}{Vidur Parkash}
 received his B.E (Hons.) degree in Electronics and Communication Engineering from Maharshi Dayanand University,
India in 2004. He is currently working towards his Ph.D. degree in Electrical Engineering at Michigan Technological University.
His research interests include computational nanoelectronics, optoelectronic devices and VLSI design.
\end{IEEEbiographynophoto}

\begin{IEEEbiographynophoto}{Anand K Kulkarni}
 received his B.Sc. and M.Sc. degrees in Physics from Karnatak University, Dharawad, India in 1967
 and 1970 respectively. He obtained M.S. in Physics(Solid State Physics) from Iowa
 State University, Ames, Iowa in 1975 and Ph.D. in Engineering(Electrical) from University of Nebraska
, Lincoln, Nebraska in 1979. He joined Michigan Technological University, Houghton, MI in 1978 as
 a visiting instructor and is currently an associate professor in the department of Electrical and
 Computer Engineering. His research interests are in electronic and photonic
 materials and devices.
\end{IEEEbiographynophoto}



\vfill


\end{document}